\documentclass[prl,twocolumn]{revtex4}%
\usepackage{amsfonts}
\usepackage{amsmath}
\usepackage{amssymb}
\usepackage{graphicx}%
\providecommand{\U}[1]{\protect\rule{.1in}{.1in}}

\begin{document}
\title{Journey Beyond the Schwarzschild Black Hole Singularity}
\author{Ignacio J. Araya, Itzhak Bars and Albin James}
\affiliation{Department of Physics and Astronomy, University of Southern California, Los
Angeles, CA 90089-0484, USA}
\keywords{Black hole, geodesic completeness}
\pacs{PACS number}

\begin{abstract}
We present the geodesical completion of the Schwarzschild black hole in four
dimensions which covers the entire space in $\left(  u,v\right)  $
Kruskal-Szekeres coordinates, including the spacetime behind the black and
white hole singularities. The gravitational constant switches sign abruptly at
the singularity, thus we interpret the other side of the singularity as a
\ region of antigravity. The presence of such sign flips is a prediction of
local (Weyl) scale invariant geodesically complete spacetimes which improve
classical general relativity and string theory. We compute the geodesics for
our new black hole and show that all geodesics of a test particle are
complete. Hence, an ideal observer, that starts its journey in the usual space
of gravity, can reach the other side of the singularity in a finite amount of
proper time. As usual, an observer outside of the horizon cannot verify that
such phenomena exist. However, the fact that there exist proper observers that
can see this, is of fundamental significance for the construction of the
correct theory and the interpretation of phenomena pertaining to black holes
and cosmology close to and beyond the singularities.

\end{abstract}
\maketitle

The Schwarzschild black hole is a spherically symmetric solution to the vacuum
Einstein equations $R_{\mu\nu}\left(  g\right)  =0$
\begin{equation}
ds^{2}=-\left(  1-\frac{r_{0}}{r}\right)  dt^{2}+\left(  1-\frac{r_{0}}%
{r}\right)  ^{-1}dr^{2}+r^{2}d\Omega^{2}. \label{bh1}%
\end{equation}
Here, $r_{0}\equiv2G_{N}M,$ is the radius of the horizon, $G_{N}$ is the
gravitational constant, and $M$ is the ADM mass of the black hole. Although
the Ricci and scalar curvatures are zero, the curvature tensor $R_{\mu
\nu\lambda\sigma}$ blows up at the $r=0$ singularity. The spacetime is better
described in terms of the Kruskal-Szekeres coordinates,
\begin{equation}%
\begin{array}
[c]{c}%
u=\pm\left\vert 1-\frac{r}{r_{0}}\right\vert ^{\frac{1}{2}}e^{\left(
r+t\right)  /2r_{0}},\\
v=\pm\left\vert 1-\frac{r}{r_{0}}\right\vert ^{\frac{1}{2}}\text{Sign}\left(
1-\frac{r}{r_{0}}\right)  e^{\left(  r-t\right)  /2r_{0}},
\end{array}
\end{equation}
that satisfy the following properties ($+$ corresponds to regions I\&II and
$-$ to regions III\&IV in Fig.1)
\begin{equation}%
\begin{array}
[c]{c}%
\text{when }uv<1\text{ or }r>0,\\
uv=\left(  1-\frac{r}{r_{0}}\right)  e^{r/r_{0}},\;\;\frac{u}{v}%
=\text{Sign}\left(  1-\frac{r}{r_{0}}\right)  e^{t/r_{0}},\\
r=r_{0}R_{+}\left(  uv\right)  ,\;t=r_{0}\ln\left\vert \frac{u}{v}\right\vert
,\\
R_{+}\left(  uv\right)  \equiv1+\text{ProductLog}[0,\frac{-uv}{e}],\\
ds^{2}=r_{0}^{2}\left(  2\frac{e^{-R_{+}\left(  uv\right)  }}{R_{+}\left(
uv\right)  }\left(  -2dudv\right)  +R_{+}^{2}\left(  uv\right)  d\Omega
^{2}\right)  .
\end{array}
\label{uv<1}%
\end{equation}

The spacetime $uv<1$ is geodesically incomplete because an observer that
starts a journey in region I on a geodesic that crosses the horizon into
region II, reaches the black hole singularity in a finite amount of
\textit{proper time}. Such geodesics are artificially stopped at the $r=0$
singularity because in conventional general relativity it is assumed there
does not exist a spacetime beyond this point. This indicates that the theory
is incomplete since it cannot answer the question of what happens as proper
time continues to tick. The geodesic incompleteness is a general problem that
occurs at every gravitational singularity, not only at the Schwarzschild black
hole.
\begin{center}
\includegraphics[
height=1.5738in,
width=1.3347in
]%
{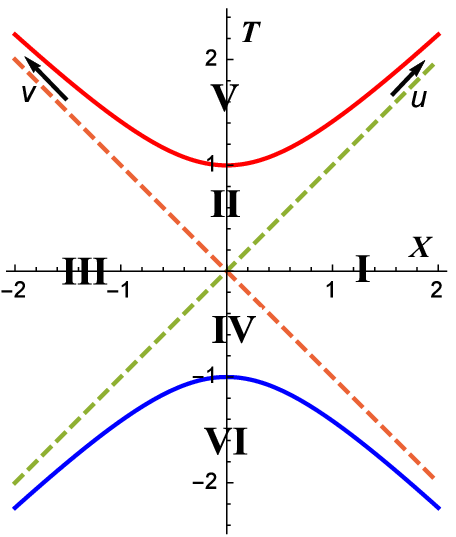}%
\\
Fig.[1]-Kruskal diagram
\end{center}
The problem occurs both in the Einstein frame and in the string frame of
general relativity, as well as in string theory when it uses geodesically
incomplete background geometries in the worldsheet formulation (there would be
incomplete string solutions similar to particle geodesics). Usually an appeal
is made to \textquotedblleft quantum gravity\textquotedblright, such as string
theory, to resolve the problems at singularities. However if the worldsheet
formalism of strings is already geodesically incomplete we may expect that
such an incomplete version of string theory would also be subject to similar
problems in its classical and quantum versions. Therefore we believe that a
healthier approach is to first understand and improve the geodesic
completeness of gravitational and string theories at the classical level and
then study the quantum effects using the improved theories. Later in this
paper we will connect to the general approach \cite{BST-conf}\cite{BST-string}
to construct geodesically complete gravitational and string theories.

We propose a geodesic completion of the Schwarzschild blackhole geometry as
follows. We note that there is another solution of $R_{\mu\nu}\left(
g\right)  =0$ which looks just like Eq.(\ref{bh1}) except for replacing
$r_{0}$ by $-r_{0}.$ This solution has no horizon and corresponds to a bare
singularity and is usually discarded. We propose a new interpretation of this
solution. We attribute the flip of sign of $r_{0}$ to be due to the flip of
sign of the gravitational constant exactly at the singularity, $G_{N}%
\rightarrow-G_{N}.$ This flip naturally occurs in general relativity
interacting with matter improved with local scale (Weyl) invariance
\cite{BST-conf} as well as in the related improved string theory
\cite{BST-string}. We therefore suggest, consistently with \cite{BST-conf}%
\cite{BST-string}, that this solution must belong to regions $V$ and $VI$ that
are behind the black or white hole singularities, and claim that those are
antigravity regions where gravity is repulsive rather than attractive. We will
show that the gravity regions ($I,II,III,IV)$ and the antigravity regions
($V,VI$) are geodesically connected by exhibiting the metric $g_{\mu\nu}$ that
spans the union of all regions and by displaying the complete set of geodesics
in this geometry that go through the black/white hole singularity. To begin,
we use new Kruskal-Szekeres coordinates to rewrite the solution
\begin{equation}
ds^{2}=-\left[  -\left(  1+\frac{r_{0}}{r}\right)  dt^{2}+\left(
1+\frac{r_{0}}{r}\right)  ^{-1}dr^{2}+r^{2}d\Omega^{2}\right]  . \label{newg}%
\end{equation}
The unusual overall minus sign is needed for the continuity of the metrics in
Eqs.(\ref{bh1},\ref{newg}) at $r=0.$ The extra sign is typical of antigravity
as explained in Eq.(\ref{pmg}). See the discussion following Eq.(\ref{pmg}) to
address any concerns about ghosts. So, in regions $V\&VI,$%
\begin{equation}%
\begin{array}
[c]{c}%
\text{when }uv>1,\;\text{and }r>0,\\
u=\pm\left(  1+\frac{r}{r_{0}}\right)  ^{-\frac{1}{2}}e^{\left(  r+t\right)
/2r_{0}},\;\;\\
v=\pm\left(  1+\frac{r}{r_{0}}\right)  ^{-\frac{1}{2}}e^{\left(  r-t\right)
/2r_{0}},\\
uv=\left(  1+\frac{r}{r_{0}}\right)  ^{-1}e^{r/r_{0}},\;\frac{u}{v}%
=e^{t/r_{0}},\\
r=r_{0}R_{-}\left(  uv\right)  ,\;t=r_{0}\ln\left\vert \frac{u}{v}\right\vert
,\\
R_{-}\left(  uv\right)  =\left(  -1-\text{ProductLog}[-1\text{,}\frac{-1}%
{euv}\text{]}\right)  ,\\
\frac{ds^{2}}{r_{0}^{2}}=2\frac{e^{-R_{-}}}{R_{-}}\left(  1+R_{-}\right)
^{2}\left(  -2dudv\right)  -R_{-}^{2}d\Omega^{2}.
\end{array}
\label{uv>1}%
\end{equation}
The function ProductLog$[k$,$z$] corresponds to branches of the Lambert
function in the complex plane. For $k=-1$ our $R_{-}\left(  uv\right)  $ is
always real and positive when $uv>1$.

The union of the $uv\lessgtr1$ regions is the metric%
\begin{equation}%
\begin{array}
[c]{c}%
ds^{2}=r_{0}^{2}\left[  -2sR^{\prime}\left(  uv\right)  \left(  -2dudv\right)
+sR^{2}\left(  uv\right)  d\Omega^{2}\right]  ,\\
R\left(  uv\right)  \equiv s\left(  1+\text{ProductLog}\left[  \frac{s-1}%
{2}\text{,}\frac{\left(  -uv\right)  ^{s}}{e}\right]  \right)  ,\\
s\equiv\text{Sign}\left(  1-uv\right)  .
\end{array}
\label{FullMetric}%
\end{equation}
In this metric \textit{space and time, (X,T) in Fig.1, do not switch roles on
the other side of the singularity. }We will show that this is a geodesically
complete spacetime of gravity and antigravity regions, noting that the
effective gravitational constant switches sign at $uv=1$
\begin{equation}
G_{N}~\text{Sign}\left(  1-uv\right)  . \label{GN}%
\end{equation}
The expression for $R\left(  uv\right)  $ given in Eq.(\ref{FullMetric}) is
the unique \textit{real} and \textit{positive} $R$ solution for the following
equation
\begin{equation}
\left(  1-sR\right)  ^{s}e^{R}=uv. \label{Ruvs}%
\end{equation}
A plot of $R\left(  uv\right)  $ and its derivative $R^{\prime}\left(
uv\right)  $ is given in Fig.(2), showing that $R$ is positive for all $uv,$
vanishes at the singularity $uv=1,$ and approaches an infinite slope
$R^{\prime}\rightarrow\pm\infty$ at that point.
\begin{center}
\includegraphics[
height=1.5519in,
width=2.5007in
]%
{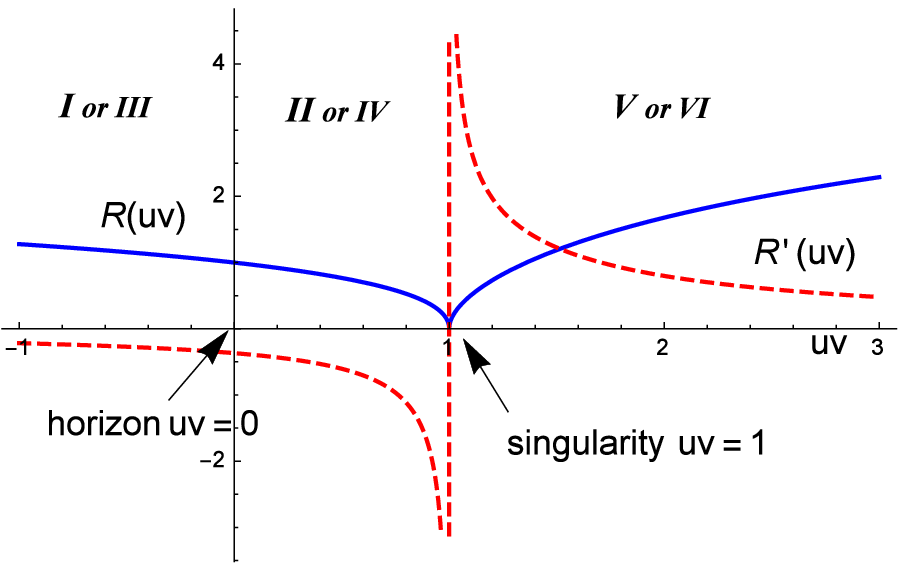}%
\\
Fig.[2] - $R\left(  uv\right)  $ solid, $R^{\prime}\left(  uv\right)  $
dashed.
\end{center}
Generally Eq.(\ref{Ruvs}) has many solutions in the complex $R$ plane. These
are expressed in terms of the many branches of the well documented function
ProductLog$\left[  k\text{,}z\right]  $. The branch in Eq.(\ref{FullMetric})
is real and positive for all real values of $uv$. We now show that the
derivative $R^{\prime}\left(  uv\right)  $ for all $uv,$ including $uv=1,$ is
given by
\begin{equation}
sR^{\prime}=\frac{e^{-R}}{-R}\left(  1-sR\right)  ^{1-s}=\frac{1-sR}{-uv~R}.
\label{R'}%
\end{equation}
With this form of $R^{\prime}$ the generalized metric in Eq.(\ref{FullMetric})
agrees with the metrics in Eqs.(\ref{uv<1},\ref{uv>1}) for $uv<1$ and $uv>1,$
and also defines the new metric at the singularity $uv=1.$ Some care is needed
to verify Eq.(\ref{R'}) since the derivative of $s$ is a delta function
$s^{\prime}=-2\delta\left(  1-uv\right)  .$ The simplest approach is to
formally take the derivative of both sides of Eq.(\ref{Ruvs}) for any
$R\left(  uv\right)  $ and $s\left(  uv\right)  ,$%
\begin{equation}
\left[
\begin{array}
[c]{c}%
\frac{1-s^{2}-sR}{1-sR}R^{\prime}\\
+\left(  \ln\left(  1-sR\right)  -\frac{sR}{1-sR}\right)  s^{\prime}%
\end{array}
\right]  \left(  1-sR\right)  ^{s}e^{R}=1. \label{verify}%
\end{equation}
$\allowbreak$Near the singularity, for small $R$ and well behaved $s,$ the
coefficient of $s^{\prime}$ in Eq.(\ref{verify}) has an expansion such that
the $s^{\prime}$ term becomes, $\left(  s^{2}\right)  ^{\prime}\left(
-R\right)  \left(  1+\frac{3}{4}sR+O\left(  \left(  sR\right)  ^{2}\right)
\right)  .$ With our $s\left(  uv\right)  $ Eq.(\ref{FullMetric}), and
$\left(  s^{2}\right)  ^{\prime}R\left(  uv\right)  \rightarrow0,$ the
$s^{\prime}$ term in Eq.$\left(  \text{\ref{verify}}\right)  $ vanishes for
all $uv,$ including $uv=1.$ The remaining $R^{\prime}$ term in
Eq.(\ref{verify}) shows that $R^{\prime}\left(  uv\right)  $ is given by
Eq.(\ref{R'}) for all $uv.$ We record here the behavior of $R$ and $R^{\prime
}$ near the singularity and far away from it%
\begin{equation}%
\begin{array}
[c]{l}%
uv\simeq1:\;\left\{
\begin{array}
[c]{l}%
R\simeq(2\left\vert 1-uv\right\vert )^{1/2},\\
sR^{\prime}\simeq-(2\left\vert 1-uv\right\vert )^{-1/2}.
\end{array}
\right. \\
\left\vert uv\right\vert \simeq\infty:\;\left\{
\begin{array}
[c]{l}%
R\simeq\ln\left(  \left\vert uv\right\vert \left(  \ln\left\vert uv\right\vert
\right)  ^{\text{Sign}\left(  uv\right)  }\right)  .\\
sR^{\prime}\simeq\frac{1}{\left\vert uv\right\vert }+\frac{1}{uv\ln\left\vert
uv\right\vert }.
\end{array}
\right.
\end{array}
\label{approx}%
\end{equation}
noting that these are consistent with the plots in Fig.(2).

Remarkably, the metric in Eq.(\ref{FullMetric}) is a solution of the vacuum
Einstein equations, $R_{\mu\nu}\left(  g\left(  u,v,\Omega\right)  \right)
=0,$ for all $\left(  u,v\right)  $. By construction, we knew that we have a
solution when $uv\neq1.$ We remark that for a metric of the form
(\ref{FullMetric}) which is fully specified by a single function $R\left(
uv\right)  $, the Ricci tensor vanishes automatically for any $R\left(
uv\right)  .$ In our case, for the specific form of $R\left(  uv\right)  $
given in Eqs.(\ref{FullMetric},\ref{R'}), we obtain $R_{\mu\nu}=0$ for all
$\left(  u,v\right)  $ including at the black and white hole singularities at
$uv=1$.

To study the geodesics we now consider a test particle of mass $m$ moving in
this improved black hole background. The worldline Lagrangian has the form
$\mathcal{L}=\frac{1}{2e}g_{\mu\nu}\left(  x\right)  \dot{x}^{\mu}\dot{x}%
^{\nu}-e\frac{m^{2}}{2},$ where $e\left(  \tau\right)  $ is the einbein. The
constraint due to $\tau$-reparametrization is the equation of motion with
respect to $e\left(  \tau\right)  .$ After choosing the gauge $e\left(
\tau\right)  =r_{0}^{2},$ which corresponds to interpreting $\tau$ as a
dimensionless \textit{proper time}, the constraint takes the form
\begin{equation}
\left[  -2sR^{\prime}\left(  uv\right)  \left(  -2\dot{u}\dot{v}\right)
+sR^{2}\left(  uv\right)  \dot{\Omega}^{2}\right]  +m^{2}r_{0}^{2}=0.
\label{Constraint}%
\end{equation}
This is the $g^{\mu\nu}p_{\mu}p_{\nu}+m^{2}=0$ constraint for our new extended
black hole metric. The canonical conjugate to the solid angle $\vec{\Omega}$
(a unit vector) is related to angular momentum $\vec{L},$ which is conserved
due to rotational symmetry in the Lagrangian or the metric (\ref{FullMetric}).
Furthermore, there is also a symmetry under opposite global rescalings of
$\left(  u,v\right)  \rightarrow\left(  \lambda u,\lambda^{-1}v\right)  $.
This amounts to translations of the time coordinate, $t/r_{0}\rightarrow
\left(  t/r_{0}+\ln\lambda^{2}\right)  ,$ as seen from Eqs.(\ref{uv<1}%
,\ref{uv>1}). Hence, there is an additional conserved quantity, that amounts
to the canonical conjugate to $t/r_{0}$, which is (up to a rescaling by
$r_{0}$) a dimensionless energy parameter $E$. Taking the conserved quantities
$\left(  E,\vec{L}\right)  $ into account, we rewrite the constraint in
Eq.(\ref{Constraint}) as follows (see derivation below)%
\begin{equation}
s\left(  \frac{-E^{2}}{uvR^{\prime}}+\frac{R^{\prime}\left(  \partial_{\tau
}\left(  uv\right)  \right)  ^{2}}{uv}+\frac{\vec{L}^{2}}{R^{2}}\right)
+m^{2}r_{0}^{2}=0. \label{A}%
\end{equation}
This equation now involves a single time-dependent degree of freedom, namely
$\left(  uv\right)  \left(  \tau\right)  ,$ whose solution as a function of
proper time $\tau$ would determine all geodesics. The second order equations
derived from the worldline Lagrangian (geodesics) are automatically solved by
the solutions of this first order differential equation because they must obey
the constraint (\ref{A}). Hence this determines \textit{all geodesics for all
possible initial conditions} $\left(  E,\vec{L}\right)  $ for a test particle
of small mass $m$.

To see how Eq.(\ref{A}) is derived from the constraint Eq.(\ref{Constraint}),
we need to clearly identify the canonical conjugate to $t\left(  \tau\right)
.$ For this purpose it is useful to transform to yet another set of
coordinates $\left(  \rho,t\right)  $ instead of $\left(  u,v\right)  $ that
still cover the entire $\left(  u,v\right)  $ plane. We leave $t$ unchanged as
in Eqs.(\ref{uv<1},\ref{uv>1}) and introduce, $\rho=1-uv,$ in the range
$-\infty<\rho<\infty.$ Thus, the coordinate transformation and its inverse is%
\[%
\begin{array}
[c]{c}%
\rho=1-uv,\;\;\frac{t}{r_{0}}=\ln\left\vert \frac{u}{v}\right\vert
\equiv\tilde{t},\\
u=\pm\sqrt{\left\vert 1-\rho\right\vert }e^{\frac{t}{2r_{0}}},\;v=\pm
\text{Sign}\left(  1-\rho\right)  \sqrt{\left\vert 1-\rho\right\vert
}e^{-\frac{t}{2r_{0}}}.
\end{array}
\]
After this change of coordinates the metric (\ref{FullMetric}) becomes%
\[
ds^{2}=r_{0}^{2}\left[  -sR^{\prime}\left(  -d\tilde{t}^{2}\left(
\rho-1\right)  +d\rho^{2}\left(  \rho-1\right)  ^{-1}\right)  +sR^{2}%
d\Omega^{2}\right]  .
\]
From the corresponding worldline Lagrangian we compute, $E,$ the canonical
conjugate to $\tilde{t}\left(  \tau\right)  ,$ and express the conserved
angular momentum, $\vec{L},$ in terms of the angular velocity $\partial_{\tau
}\vec{\Omega}\left(  \tau\right)  $
\begin{equation}
E=\left(  \rho-1\right)  sR^{\prime}\partial_{\tau}\tilde{t},\;\;\vec
{L}=sR^{2}\partial_{\tau}\vec{\Omega}.
\end{equation}
After rewriting $\left(  \dot{u},\dot{v},\dot{\Omega}\right)  $ in terms of
$\left(  \dot{t},\dot{\rho},\vec{L}\right)  ,$ the constraint
(\ref{Constraint}) takes the form of Eq.(\ref{A}).

Now, it is easy to get an intuitive understanding of the time development of
$\left(  uv\right)  \left(  \tau\right)  ,$ or equivalently $\rho\left(
\tau\right)  =1-\left(  uv\right)  \left(  \tau\right)  ,$ by rewriting the
constraint (\ref{A}) in the form of a non-relativistic \textquotedblleft
Hamiltonian\textquotedblright\ $\mathcal{H}$ (i.e. kinetic energy + potential
energy) for one degree of freedom, subject to the condition that the
corresponding \textquotedblleft energy\textquotedblright\ level is zero,
namely $\mathcal{H}=0$ (the constraint), as follows
\begin{equation}
\mathcal{H}\equiv\frac{1}{2}\left(  \partial_{\tau}\left(  uv\right)  \right)
^{2}+V\left(  uv\right)  =0.\label{h}%
\end{equation}
This exercise identifies the potential $V\left(  uv\right)  $
\begin{equation}
V\left(  uv\right)  \equiv\frac{uv}{2R^{\prime}}\left(  \frac{\vec{L}^{2}%
}{R^{2}}+sm^{2}r_{0}^{2}\right)  -\frac{E^{2}}{2R^{\prime2}}.\label{V}%
\end{equation}
Plots of $V\left(  uv\right)  $ for small $\left\vert m^{2}\right\vert
r_{0}^{2}$ are given in Figs.(3,4).
\begin{center}
\includegraphics[
height=1.5509in,
width=2.4688in
]%
{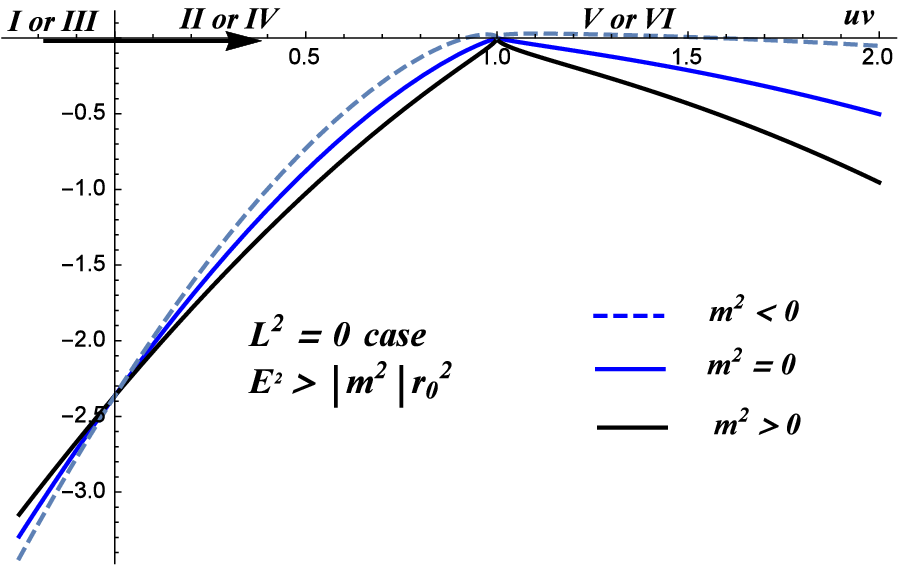}%
\\
Fig.[3] - $V\left(  uv\right)  $ for $L=0.$ Middle curve for $m=0.$%
\end{center}
\begin{center}
\includegraphics[
height=1.6559in,
width=2.6676in
]%
{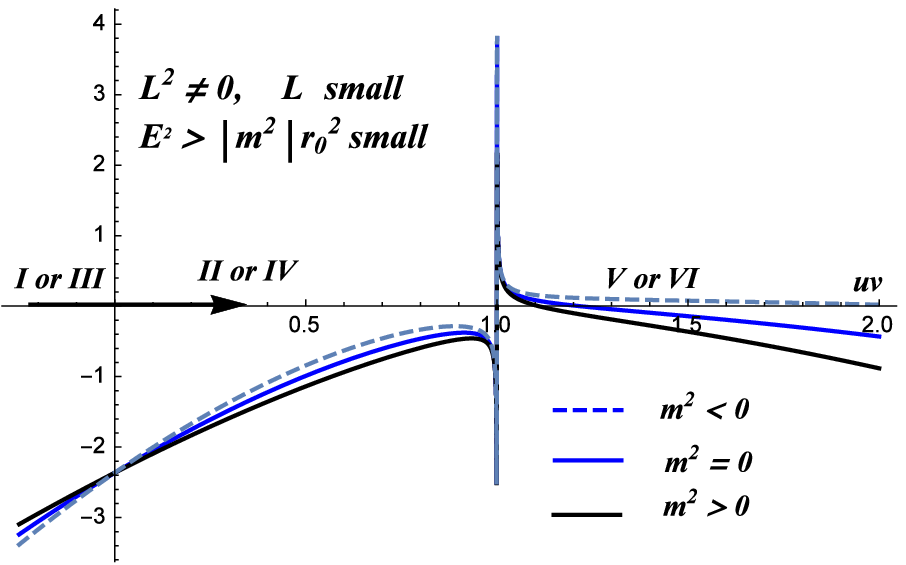}%
\\
Fig.[4] - $V\left(  uv\right)  $ for~$L\neq0.$ Middle curve for $m=0.$%
\end{center}
The features of the plots follow from the approximate behavior near to and far
from the singularity \ \ \ \ \ \ \ \ \
\[%
\begin{array}
[c]{l}%
uv\simeq1:V\simeq-\left[  \frac{uv~\left(  ~s\vec{L}^{2}+2\left\vert
1-uv\right\vert ~m^{2}r_{0}^{2}\right)  }{2\sqrt{2}\sqrt{\left\vert
1-uv\right\vert }}+E^{2}\left\vert 1-uv\right\vert \right]  ,\\
uv\simeq\pm\infty:V\simeq\frac{\left(  uv\right)  ^{2}}{2}\left[
-E^{2}+sm^{2}r_{0}^{2}+\frac{L^{2}}{\left(  \ln\left\vert uv\right\vert
\right)  ^{2}}\right]  .
\end{array}
\]
In this potential, the dimensionless parameters $(E^{2},\vec{L}^{2})$ may be
considered as initial conditions for the particle of dimensionless mass
$m^{2}r_{0}^{2}$. We omit the discussion of particles trapped in orbits around
the black hole, that would occur when $E^{2}<m^{2}r_{0}^{2},$ as this does not
change our main points. Then, the asymptotic form of Eq.(\ref{A}) for large
$\left\vert uv\right\vert $ is, $-E^{2}+\vec{p}^{2}+m^{2}r_{0}^{2}=0,$
indicating that $E$ must satisfy $E^{2}>m^{2}r_{0}^{2}$. We consider small
values of $\left\vert m^{2}r_{0}^{2}\right\vert $ since huge masses would
violate the spirit that $m$ represents a small probe for which the back
reaction of the black hole can be neglected.

The constraint (\ref{h}) is equivalent to a first order differential equation,
$\partial_{\tau}\left(  uv\right)  =\pm\sqrt{-2V\left(  uv\right)  },$ for the
single variable $uv$ whose solution is
\begin{equation}
\left(  \tau-\tau_{0}\right)  =\pm\int_{u_{0}v_{0}}^{uv}dx/\sqrt{-2V\left(
x\right)  }, \label{tau}%
\end{equation}
where the $\pm$ signs are chosen according to whether the initial velocity is
toward or away from the black hole. This expression can in principle be solved
for $uv$ as a function of $\tau,$ yielding the desired solution $\left(
uv\right)  \left(  \tau\right)  =F\left(  \tau\right)  $ where $F\left(
\tau\right)  $ is fully determined. Although this appears to be complicated we
note that from the plots of $V\left(  uv\right)  $ alone we can easily obtain
an intuitive feeling of all possible motions that $\left(  uv\right)  \left(
\tau\right)  $ can perform.

Consider the case of zero angular momentum ($\vec{L}=0,$ Fig.3). A particle
that obeys the constraint in Eq.(\ref{h}) has the same $\mathcal{H}$-energy
level as the summit of the \textquotedblleft mountain\textquotedblright. Such
a particle that comes from region $I$ where $uv<0$ (approaching from the left
in Fig.3) will keep climbing the $V\left(  uv\right)  $ mountain, passing into
region $II$ at the horizon at $uv=0$, and (in the case of $m^{2}\geq0$)
reaching the peak of the mountain where it slows down and stops momentarily at
the $uv=1$ singularity. Note that at the peak the potential vanishes, $\left.
V\left(  uv=1\right)  \right\vert _{L=0}=0$, and therefore $\partial_{\tau
}\left(  uv\right)  =0$ to satisfy the constraint (\ref{h}). In fact, at the
singularity, $\dot{u}=\dot{v}=0$ when we examine the rest of the equations of
motion that follow from the worldline Lagrangian $\mathcal{L}$, so the
particle stops temporarily at the summit of the mountain in Fig.3. This
journey takes a finite amount of proper time $\tau$ because the integral in
Eq.(\ref{tau}) is finite. Clearly the summit is an unstable point, so at the
subsequent moment in proper time, the particle will either slide forward to
$uv>1$ down the mountain into region $V$ where there is antigravity, or slide
back to $uv<1$ into region $II$ and then $III$ where there is gravity. It will
not slide back into regions $II$ and then $I$ because this would cause closed
timelike curves, and indeed this can be deduced from analytic investigations.
Forward $uv$ or backward $uv$ at $uv=1$ are allowed solutions of the geodesic
equations of motion, so both of them will happen. In either case \textit{the
particle moves on to another world} that is geodesically connected to the
original starting point in region $I.$ This shows that particles that fall
into the black hole (beyond the horizon) will inevitably end up in a new
universe according to this classical analysis. Note that gravity observers in
region $III$ will interpret that the particle comes out of a white hole, while
those in the antigravity region $V$ will interpret it as coming out of a naked singularity.

A quantum analysis that treats a small probe in a static black hole (as in the
present case) will reach the same conclusion and provide non-vanishing
probability amplitudes for transmission to regions $III$ and $V$ (for tachyons
as well). The computation can be performed in a WKB approximation just as in
non-relativistic potential theory as presented elsewhere.

Next we consider non-zero angular momentum ($L^{2}\neq0,\;$Fig.4). In this
case the particle coming from regions $I\&II$ hits an angular momentum barrier
at $uv=1,$ so classically it can only bounce back to regions $II\&III.$
However, quantum mechanically there will be a non-zero transmission
probability to also tunnel into region $V.$

To be fully convinced of our intuitive analysis it is useful to have some
analytic expressions for the geodesics. This looks complicated in 4-dimensions
although there is no problem numerically. However, for the closely related
2-dimensional stringy black hole, after geodesically completing its space-time
as we did above, we have explicitly constructed the full set of geodesics of
the type $L=0$ discussed above. This classical result fully supports the
intuitive discussion. This work, together with the corresponding quantum
computation, will be reported in a separate publication.

In a similar way we can also discuss the geodesics whose initial conditions
begin in the regions $V$ or $VI$.

We now give a short description of how our geodesically complete black hole
spacetime in Eq.(\ref{FullMetric}) fits perfectly with the Weyl symmetric
re-formulation of geodesically complete gravity (SM+GR) \cite{BST-conf} and
string theory (ST) \cite{BST-string}. We concentrate only on the basic
consequence of the Weyl symmetry, which is that dimensionful parameters are
not allowed. All dimensionful constants of phenomenological significance,
including the Newton constant (and therefore the string tension) emerge from
Weyl-gauge fixing of some gauge degrees of freedom \cite{BST-conf}%
\cite{BST-string}. As an illustration, consider the case of the SM+GR which
contains the SU$\left(  2\right)  \times$U$\left(  1\right)  $ Higgs doublet
$H$ and an additional singlet scalar $\phi$ required by the Weyl-symmetric
approach. $\phi$ is compensated by the Weyl symmetry, so $\phi$ is not a true
additional degree of freedom, but participates in an important structure of
the symmetry that has physical consequences. Due to the symmetry all scalars
are \textquotedblleft conformally coupled\textquotedblright, implying the
special non-minimal coupling to the curvature
\begin{equation}
\frac{1}{12}\left(  \phi^{2}-s^{2}\right)  R\left(  g\right)  ,\;\text{with
}s^{2}\equiv2H^{\dagger}H. \label{signs}%
\end{equation}
This structure is the same in low energy ST \cite{BST-string} but with a
different interpretation of $s$. The relative minus sign in Eq.(\ref{signs})
is obligatory. Weyl symmetry requires that, with the signs above, $\phi$ has
the wrong sign kinetic energy while $H$ has the correct sign. So, $\phi$ is a
ghost but, since it can be removed by a Weyl gauge choice, this is not a
problem. If $\phi$ were not a ghost then the curvature term would have a
purely negative coefficient, $-\frac{1}{12}\left(  \phi^{2}+s^{2}\right)  ,$
which leads to only a purely negative gravitational constant, so there are no
alternatives to (\ref{signs}). Therefore, the effective Planck mass $\frac
{1}{12}\left(  \phi^{2}-s^{2}\right)  $ (or the Newton constant) is not
positive definite. At the outset of this approach in 2008 the immediate
question was whether the dynamics would allow $\left(  \phi^{2}-s^{2}\right)
$ to remain always positive. It was eventually determined in 2010-2011
(references in \cite{BST-conf}\cite{BST-string}) that the solutions of the
field equations that do not switch sign for this quantity are non-generic and
of measure zero in the phase space of initial conditions for the fields
$\left(  \phi,s\right)  $. So, according to the dynamics, it is untenable to
insist on a limited patch of field space. By contrast, it was found that the
theory becomes geodesically complete when all field configurations are
included, thus solving generally the basic problem of geodesic incompleteness.

With a gauge choice, extra Weyl gauge degrees of freedom can be removed, but
one can err by choosing an illegitimate gauge that corresponds to a
geodesically incomplete patch. Indeed this is what happens in the
\textquotedblleft Einstein gauge\textquotedblright\ (E) and in the
\textquotedblleft string gauge\textquotedblright\ (s)\ \cite{BST-conf}%
\cite{BST-string}. \
\[%
\begin{array}
[c]{l}%
\text{E-gauge:\ }\frac{1}{12}\left(  \phi_{E+}^{2}-s_{E+}^{2}\right)
=\frac{+1}{16\pi G_{N}},\\
\text{s-gauge:\ }\frac{d-2}{8\left(  d-1\right)  }\left(  \phi_{s+}^{2}%
-s_{s+}^{2}\right)  =\frac{+1}{2\kappa_{d}^{2}}e^{-2\Phi},\;\Phi
=\text{dilaton.}%
\end{array}
\]
Conventional general relativity and string theory are geodesically incomplete
because the gauge choices just shown are valid only in the field patch in
which $\left\vert \phi\right\vert >\left\vert s\right\vert $. The dynamics
contradict the assumption of gauge fixing to only the positive patch. In the
negative regions one may choose again the Einstein or string gauge, but now
with a negative gravitational constant, $\frac{1}{12}\left(  \phi_{E-}%
^{2}-s_{E-}^{2}\right)  =\frac{-1}{16\pi G_{N}},$ or $\frac{d-2}{8\left(
d-1\right)  }\left(  \phi_{s-}^{2}-s_{s-}^{2}\right)  =\frac{-1}{2\kappa
_{d}^{2}}e^{-2\Phi}$; in those spacetime regions gravity is repulsive
(antigravity). In the corresponding worldsheet formulation of string theory
the string tension also switches sign \cite{BST-string}. Thus the Weyl
symmetric (SM+GR) or string theory predict that, in the Einstein or string
gauges, one should expect a sudden sign switch of the effective Planck mass
$\frac{1}{12}\left(  \phi^{2}-s^{2}\right)  $ at certain spacetime points that
typically correspond to singularities (e.g. big bang, black holes) encountered
in the Einstein or string frames. As shown in \cite{BST-conf}\cite{BST-string}
one may choose better Weyl gauges (e.g. \textquotedblleft$\gamma
$-gauge\textquotedblright, choose $\det\left(  -g\right)  \rightarrow1$, or
\textquotedblleft$c$-gauge\textquotedblright, choose $\phi\rightarrow
$constant$)$ that cover globally all the positive and negative patches. Then
the sign switch of the effective Planck mass $\frac{1}{12}\left(  \phi
^{2}-s^{2}\right)  $ is smooth rather than abrupt. However, if one wishes to
work in the Einstein or string frames, as we did in this paper, to recover the
geodesically complete theory one must allow for the gravitational constant to
switch sign at singularities, as in Eq.(\ref{GN}), and connect solutions for
fields across gravity/antigravity patches. In the $\pm$ Einstein gauges
Eq.(\ref{signs}) becomes
\begin{equation}
\frac{\left(  \phi_{E\pm}^{2}-s_{E\pm}^{2}\right)  R\left(  g_{E\pm}\right)
}{12}=\frac{R\left(  g_{E\pm}\right)  }{\pm16\pi G_{N}}=\frac{R\left(  \pm
g_{E\pm}\right)  }{16\pi G_{N}}. \label{pmg}%
\end{equation}
where the $\pm$ for the gravity/antigravity regions can be absorbed into a
redefinition of the signature of the metric, $\tilde{g}_{\mu\nu}^{E}=\pm
g_{\mu\nu}^{E\pm}$ \cite{BST-string}. Our new black hole in
Eq.(\ref{FullMetric}) is for the \textit{continuous} $\tilde{g}_{\mu\nu}^{E}$
in the \textit{union} of the gravity/antigravity patches. This explains the
extra minus sign in Eq.(\ref{newg}) and connects it to the underlying Weyl
symmetric theory.

One may be worried that the sign switches of the gravitational constant or the
string tension may lead to problems like unitarity or negative kinetic energy
ghosts. For example, in the SM+GR action in the $\pm$ Einstein gauge, some
terms in the antigravity sector flip sign and some don't \cite{BST-string}
when $g_{\mu\nu}^{E-}\rightarrow-g_{\mu\nu}^{E-}$ (e.g. $F_{\mu\nu}F^{\mu\nu}$
does not, but $R\left(  g\right)  $ does as in (\ref{pmg})). We should mention
that Ref.\cite{BST-string} has already settled that there are no unitarity
problems due to sign flips in field/string theories. As to the negative
kinetic energy concerns in antigravity (as in $-R_{-}^{2}\left(  uv\right)
\dot{\Omega}^{2}\,$in (\ref{uv>1})), this apparent instability is rendered
harmless by insisting that the only reasonable interpretation of the theory is
by \textit{observers in the gravity sector } (details and examples in an
upcoming paper \cite{lectures}). Such observers cannot experience the negative
kinetic energy in antigravity directly, but can only detect in and out signals
that interact with the antigravity region. This is no different than a closed
spacetime box\ for which the information about its interior is scattering
amplitudes for in/out states at its exterior. An analogous situation for a
cosmological singularity \cite{BST-conf}\cite{BST-string} is treated in detail
in \cite{turok}. So, there are no issues of fundamental principles.

We have demonstrated that the Schwarzschild blackhole has a geodesic
completion, and that proper observers can in principle travel through the
singularity. These results generalize to black holes in other dimensions as we
will demonstrate in additional papers. New avenues have just opened for
information to travel beyond the horizon and singularity. A similar result
that was first obtained for cosmological singularities has been applied to
develop a completely new perspective for the role of antigravity just before
the big bang (see references in \cite{BST-conf}\cite{BST-string}). Similarly,
our findings must have implications for our understanding of black holes, the
role of the geodesically complete spacetime in their formation and
evaporation, the information loss problem, and for investigating how new
physics beyond singularities in classical and quantum gravity/string theory
impacts observations in our own universe. We are at the beginning of a large project.

\end{document}